\documentclass[10pt]{iopart}
\usepackage{graphicx}
\usepackage{iopams}
\begin{document}
\title[Concentrated Model of a Double Barrier Memristive Device]{An Enhanced Lumped Element Electrical Model\\ of a Double Barrier Memristive Device }
\author{Enver~Solan\,$^{1,*}$, Sven~Dirkmann\,$^{2}$, Mirko~Hansen\,$^{3}$, Dietmar~Schroeder\,$^{4}$, Hermann~Kohlstedt\,$^{3}$, Martin~Ziegler\,$^{3}$, Thomas~Mussenbrock\,$^{5}$, and Karlheinz~Ochs\,$^{1}$}
\address{$^{1}$Chair of Digital Communication Systems, Department of Electrical Engineering and Information Science, Ruhr University Bochum, D-44780 Bochum, Germany\\
$^{2}$Institute of Theoretical Electrical Engineering, Department of Electrical Engineering and Information Science, Ruhr University Bochum, D-44780 Bochum, Germany\\
$^{3}$Nanoelektronik, Technische Fakult\"at, Christian-Albrechts-Universit\"at zu Kiel, D-24143 Kiel, Germany\\
$^{4}$Institute of Nano- and Medical Electronics, Hamburg University of Technology, D-21073 Hamburg, Germany\\
$^{5}$Electrodynamics and Physical Electronics Group, Brandenburg University of Technology Cottbus-Senftenberg, D-03046 Cottbus, Germany}
\ead{Enver.Solan@rub.de}
\vspace{10pt}
\begin{indented}
\item[]January 2017
\end{indented}

\begin{abstract}
The massive parallel approach of neuromorphic circuits leads to effective methods for solving complex problems. It has turned out that resistive switching devices with a continuous resistance range are potential candidates for such applications. These devices are memristive systems - nonlinear resistors with memory. They are fabricated in nanotechnology and hence parameter spread during fabrication may aggravate reproducible analyses. This issue makes simulation models of memristive devices worthwhile. 

Kinetic Monte-Carlo simulations based on a distributed model of the device can be used to understand the underlying physical and chemical phenomena. However, such simulations are very time-consuming and neither convenient for investigations of whole circuits nor for real-time applications, e.g. emulation purposes. Instead, a concentrated model of the device can be used for both fast simulations and real-time applications, respectively. We introduce an enhanced electrical model of a valence change mechanism (VCM) based double barrier memristive device (DBMD) with a continuous resistance range. This device consists of an ultra-thin memristive layer sandwiched between a tunnel barrier and a Schottky-contact. The introduced model leads to very fast simulations by using usual circuit simulation tools while maintaining physically meaningful parameters.

Kinetic Monte-Carlo simulations based on a distributed model and experimental data have been utilized as references to verify the concentrated model.
\end{abstract}

%
\vspace{2pc}
\section{Introduction}
Resistive switching devices are essential components for today's nonvolatile memory applications. In general, they are built by a capacitor-like metal-insulator-metal structure. Depending on material compositions, different chemical and physical effects lead to a change of the total resistance. It should be stressed that all resistive switching devices can also be interpreted as memristive systems~\cite{chua_resistance_2011}, which are in general nonlinear resistors with memory~\cite{chua_memristor-missing_1971,chua_memristive_1976}. Especially memristive systems with a continuous resistance range are potential candidates for neuromorphic circuits~\cite{kim_memristor_2012,pershin_neuromorphic_2012}.

For our approach, we focus on a double barrier memristive device (DBMD)~\cite{hansen_double_2015}, which consist of an ultra-thin memristive layer sandwiched between a tunnel barrier and a Schottky-like contact. Several benefits as a continuous resistance range, an intrinsic current compliance, an improved retention, no need for an electric forming procedure and low power consumption make the DBMD particularly suitable for neuromorphic circuits. Just as for other electronic components, a parameter spread especially for memristive devices fabricated in nanotechnology is unavoidable. This parameter spread aggravates reproducible analyses. A kinetic Monte-Carlo model of this device with distributed parameters can help to understand the underlying chemical and physical phenomena~\cite{dirkmann_kinetic_2015,dirkmann_role_2016}. However, investigations with simulations based on a distributed model are very time-consuming. Because of this, a distributed model is neither convenient for simulations of whole circuits nor for real-time applications, e.g. emulation purposes. 

Our intention is to build a replica of the device using an enhanced lumped element electrical (concentrated) model for fast simulations of both, the device itself and complex circuits including such devices. This approach also allows for reproducible analyses~\cite{ochs_sensitivity_2016}.

\begin{figure}[ht!]
	\centering
	\includegraphics[scale=0.5]{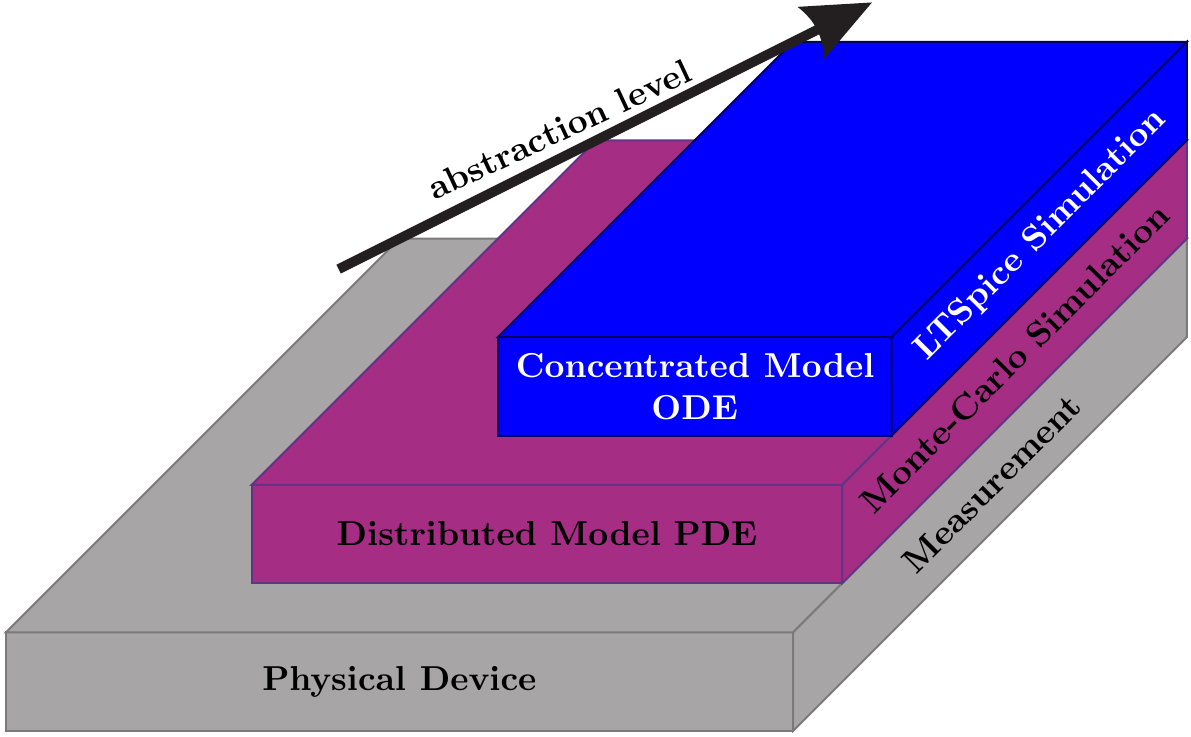}%
	\caption{Hierarchical presentation of different modeling stages from experiments through a distributed up to a concentrated model.}%
	\label{fig.:procedure}%
\end{figure}

Based on a distributed model from~\cite{dirkmann_role_2016} and on the electrical model from~\cite{hansen_double_2015} an enhanced electrical model of the DBMD with concentrated parameters has been built up and simulated using LTSpice~\cite{_linear_????}, cf. figure~\ref{fig.:procedure}.
We have verified the concentrated model by comparisons with kinetic Monte-Carlo simulations as well as measurements. Additionally, Ochs \etal\cite{ochs_wave_2016a,ochs_wave_2016b} has shown that the concentrated model can be used for emulation purposes in real-time applications.

\section{The Double Barrier Memristive Device}\label{sec:realDevice}%
\begin{figure}[!ht]
	\centering
	\includegraphics[scale=0.9]{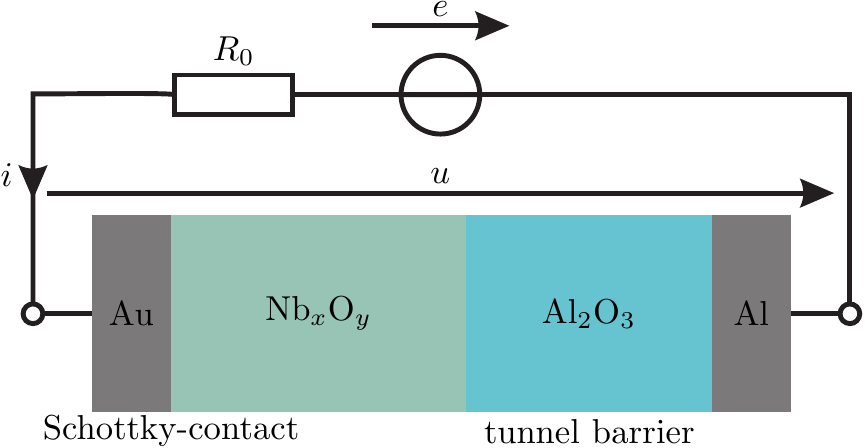}%
	\caption{Material composition of the double barrier memristive device and the measurement scenario. The externally applied voltage is $e$, with internal source resistance $R_0=0.1\:\Omega$ and the voltage drop over the device is denoted by $u$, whereas $i$ describes the current through the device.}%
	\label{fig.:realDevice}%
\end{figure}

The DBMD has been introduced by Hansen \etal\cite{hansen_double_2015}. It consists of an ultra-thin niobium oxide Nb$_x$O$_y$ followed by an aluminum oxide Al$_2$O$_3$ layer sequence sandwiched between a gold Au and an aluminum Al electrode, see figure~\ref{fig.:realDevice}.
A metal-semiconductor junction at the Au/Nb$_x$O$_y$ interface creates a Schottky-contact, whereas the Al$_2$O$_3$ layer is an electrically high quality tunnel barrier. For the DBMD the resistance change originates from oxygen diffusion caused by an externally applied electrical field resulting in modifications of local states within the Nb$_x$O$_y$ solid state electrolyte~\cite{hansen_double_2015,dirkmann_role_2016}. These modifications influence the interface properties of both the Schottky-contact and tunnel barrier, simultaneously, which in turn leads to an overall resistance change.

\subsection*{Measured Hysteresis Curve}
\begin{figure}[!ht]
	\centering
	\includegraphics[scale=0.75]{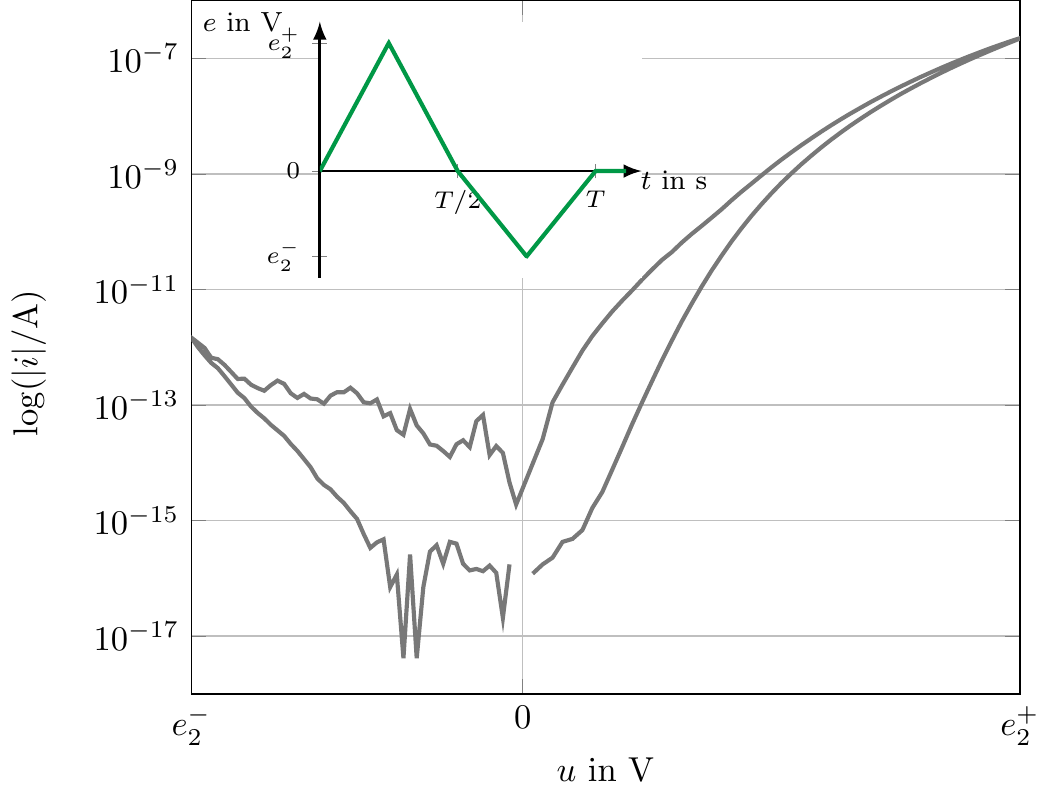}%
	\caption{Triangle-shaped input voltage (inset) and measured $i$-$u$-hysteresis curve in a logarithmic scale with respect to the absolute value of the current, normalized for a cross sectional area of $1~\mu\mathrm{m}^2$.}
	\label{fig.:measuredIV}%
\end{figure}

The DBMD has been characterized by applying a triangle-shaped voltage according to the inset of figure~\ref{fig.:measuredIV} and measuring the current, normalized for a cross sectional area of $1~\mu\mathrm{m}^2$, cf.~\cite{hansen_double_2015} and~\cite{dirkmann_role_2016}. When plotting the current versus the voltage, see figure~\ref{fig.:measuredIV}, the familiar hysteresis curve appears that is typical for memristive devices.
A logarithmic scale increases the visibility of a continuous transition between high and low resistance state. The experimentally measured data is utilized as a reference for verifying both the distributed as well as the concentrated model.

\section{Description of Device Physics}\label{sec:physic}%
A distributed model of the device is useful for a correct interpretation of physical and chemical phenomena within the device. Dirkmann \etal\cite{dirkmann_role_2016} has used such a model to investigate physical and chemical effects, which are responsible for the memristive functionality. There, investigations have been done by using a kinetic Monte-Carlo model. Based on investigations of~\cite{dirkmann_role_2016} and actual insights, we have extended the known model of Hansen \etal\cite{hansen_double_2015} to the presented concentrated model. For this progress, a qualitative recapitulation of the resistive switching behavior on the atomic level is presented in the following. A long time scale investigation with the step response of the device belongs to novel approaches within the frame of this work.

\subsection*{Resistive Switching Behavior}
Investigations of~\cite{dirkmann_role_2016} yield that the current through the device depends on the defect distribution within the memristive Nb$_x$O$_y$ layer. More precisely, the defect distribution is assumed to influence the effective thickness of the tunnel barrier as well as the Schottky-barrier height. Regarding the sign of an externally applied voltage, different physical processes occur. The following description is segmented into regimes based on the excitation of figure~\ref{fig.:measuredIV} (inset).

\subsubsection*{Thermodynamical equilibrium:}
\begin{figure}[!ht]
	\centering
	\includegraphics[scale=0.9]{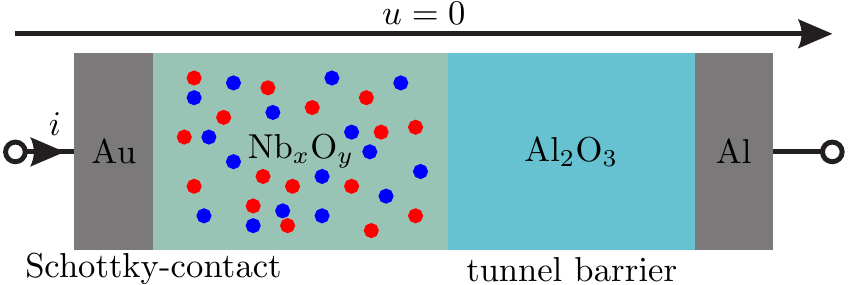}%
	\caption{Thermodynamical equilibrium originated by a homogeneous distribution of positive metal (red) and negative oxygen (blue) ions within the electrolyte.}%
	\label{fig.:physicalDescriptionDeviceThermodynamicalEq}%
\end{figure}

Without an external voltage, the device is in its thermodynamical equilibrium. Due to the Coulomb potential, positive and negative charges are uniformly distributed within the electrolyte, see figure~\ref{fig.:physicalDescriptionDeviceThermodynamicalEq}.
This is also the high resistance state of the memristive device, because of a large Schottky-barrier height and tunnel barrier thickness.

\subsubsection*{Positive applied external voltage $0<t<T/2$:}
A positive applied voltage leads to a voltage drop over the Nb$_x$O$_y$ layer. The resulting electrical field exerts a motion of charged defects. Physically, the ion motion is based on an ion-hopping phenomenon~\cite{meyer_oxide_2008}. The rate of ion motion from one stable position to the other can be described by the Arrhenius-law
\begin{equation}
k = \nu\,\rme^{-\frac{\Phi_\mathrm{a}}{k_\mathrm{B}\vartheta}}\:.
\label{eqn:arrheniusLaw}%
\end{equation} 
Here $\Phi_\mathrm{a}$ is the activation energy of defect motion within the electrolyte in J, $\vartheta$ is the temperature in K, $\nu$ is the hopping frequency in Hz and $k_\mathrm{B}$ is the Boltzmann constant in J/K.
\begin{figure}[!ht]
	\centering
	\includegraphics[scale=0.9]{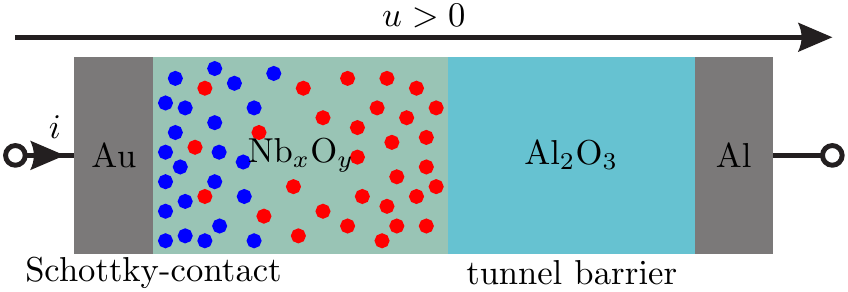}%
	\caption{A positive applied voltage results in a simultaneous enrichment of negative ions (blue) at the Schottky-contact and positive ions at the tunnel barrier. The negative ions lead to a surface potential, which decreases the Schottky-barrier height, whereas at the tunnel barrier interface a metalization procedure occurs, which in turn decreases the effective tunnel barrier thickness.}%
	\label{fig.:physicalDescriptionDevicePositiveVoltage}%
\end{figure}
The ion motion due to the applied electrical field results in an enrichment of negative ions at the Schottky-interface and positive ions at the tunnel barrier. Regarding the interface potential at the Schottky-contact, the barrier height decreases with increasing number of ions at the contact. On the other hand, the local vacancy concentration at the Al$_2$O$_3$ interface is of particular importance for the resulting electron
tunneling current. An increase of the vacancy concentration decreases the effective tunnel barrier thickness and therefore the tunneling current increases, cf. figure~\ref{fig.:physicalDescriptionDevicePositiveVoltage}. In total, the device changes gradually from a high to a low resistance state.

\subsubsection*{Negative applied external voltage $T/2<t<T$:}
\begin{figure}[!ht]
	\centering
	\includegraphics[scale=0.9]{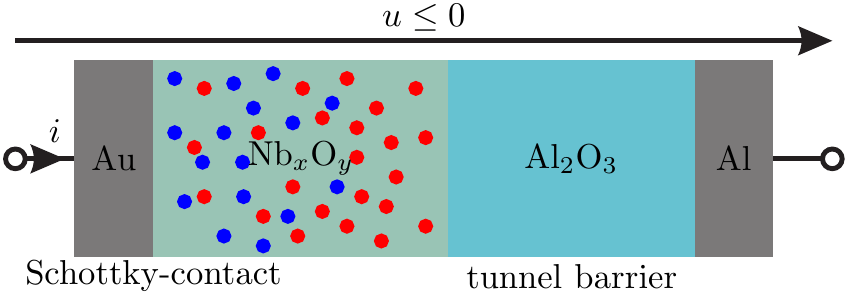}%
	\caption{For negative applied voltages, the negative oxygen ions (blue) move back into the electrolyte to restore the thermodynamical equilibrium.}%
	\label{fig.:physicalDescriptionDeviceNegativeVoltage}%
\end{figure}

For negative applied voltages, the Schottky-contact almost totally blocks the current, so that the voltage drops nearly completely across this contact. This results in an increased electric field near the interface. If the field becomes high enough, the oxygen ions are detached (desorbed) from the interface and move back into the electrolyte to finally reconstitute thermodynamical equilibrium. This phenomenon, which is the reset process, is illustrated in figure~\ref{fig.:physicalDescriptionDeviceNegativeVoltage}.
The desorption mechanism of ions at the Schottky-interface has to be taken into account in the concentrated model, because it dominates the time scale for negative applied voltages~\cite{dirkmann_role_2016}.

\subsection*{Long Time Scale Investigation}
\begin{figure}[!ht]
	\centering
	\includegraphics[scale=0.75]{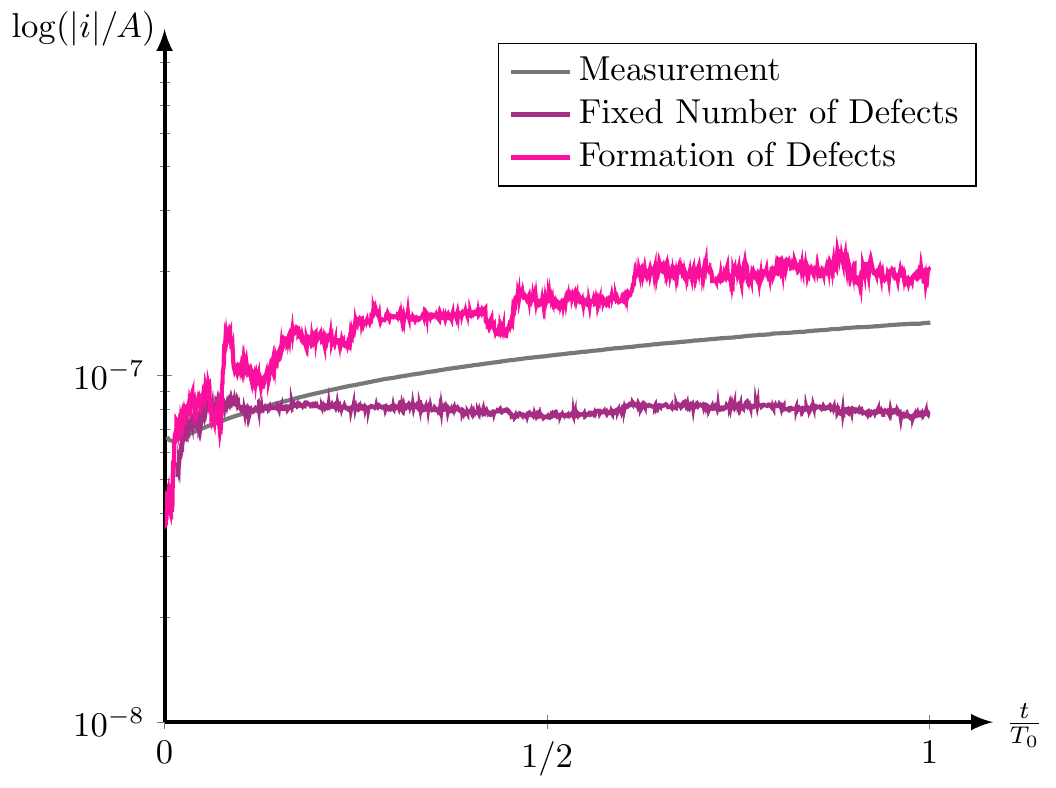}%
	\caption{Current through the DBMD for an input voltage amplitude of $2.5\:\mathrm{V}$ and $T_0=600\:\mathrm{s}$. The step responses resulting from kinetic Monte-Carlo simulations with and without defect formation are compared to the measurement.}%
	\label{fig.:Step2_5_Old}%
\end{figure}

Within the frame of this work, firstly a step response of the DBMD device was investigated. Thereby it turned out, that the assumption of a fixed number of defects within the Nb$_x$O$_y$ electrolyte might not be valid for long time scales.
In figure~\ref{fig.:Step2_5_Old} the simulated currents assuming a constant defect number and defect formation within the electrolyte are compared with the measured current. The simulation with a fixed number of defects was done using the model presented by Dirkmann \etal\cite{dirkmann_role_2016}. A constant voltage of $2.5\:\mathrm{V}$ has been chosen for the simulation as well as for the experiment. When a constant set voltage, which is sufficient to move point defects within the electrolyte, is applied to the device, the defects move towards the Au electrode. During this process, the device resistance decreases and therefore the current through the device increases, see figure~\ref{fig.:Step2_5_Old}. For longer times, all defects concentrate at the Au electrode. A further resistance change is inhibited and the current goes into saturation. In contrast to this, the measured current rises further for long time scales.
It is well-known, that a strong electric field applied at an oxide material leads to formation of point defects, as oxygen vacancies and oxygen interstitials~\cite{kwon_oxygen_2015}. Thus it stands to reason, that for long time scales a formation of point defects within the electrolyte region might also occur within the DBMD. This assumption has been tested by an extended version of the model presented in~\cite{dirkmann_role_2016} towards the defect formation. The formation of point defects is included using the rate equation~(\ref{eqn:arrheniusLaw}), see~\cite{abbaspour_role_2015,sadi_physical_2015}. The simulation result changes then with a variable number of defects within the electrolyte, cf. figure~\ref{fig.:Step2_5_Old}. The good agreement between simulation and experiment indicates a defect formation process within the DBMD for long time scales and high electric fields. It is notable, that this process is a long time process that does not affect processes on a short time scale. Although this model is able to explain the measured current for long time scales very well, it needs to be said, that it cannot be excluded that another explanation model for the current behaviour on long time scales can be found. The following kinetic Monte-Carlo simulations use a variable defect number within the electrolyte in order to verify the lumped element model.

\section{Electrical Description with Concentrated Parameters}\label{sec:concentrated}%
The distributed model leads to a better understanding of underlying physical and chemical effects. But for circuit simulations a concentrated model is more appropriate. In order to preserve the relationship to physical phenomena, we derive an enhanced electrical model with physically meaningful concentrated parameters. In the following, a derivation of the model and a classification as a memristive system are presented.

\subsection*{Memristive Systems}
The resistive switching device is in general a memristive system with an internal state. A mathematical classification of the device into the theory of general memristive systems as given by Chua \etal\cite{chua_memristive_1976} is missing. The concentrated model enables integrating the DBMD into the general memristive system theory. For this reason, we start with a brief review of these systems.

The $n$th-order voltage-controlled memristive system
\numparts
	\begin{eqnarray}
	u(t)
	&= \hat{R}(\boldsymbol{z},u,t)\,i(t)\:,  
	\label{eqn:memristiveSystemU}%
	\\
	\dot{\boldsymbol{z}}
	&=\boldsymbol{f}(\boldsymbol{z},u,t)
	\label{eqn:memristiveSystemDz}
	\end{eqnarray}
\endnumparts
is made up of an algebraic equation for the input-output relation together with a differential equation, which describes the memristive behavior or dynamics of the system~\cite{chua_memristive_1976}. Actually, the memristive system $\hat{R}$ is a generalized response, which interrelates the input voltage $u$ with the output current $i$. The current and voltage are scalars and hence $\hat{R}$ is a scalar function, which is generally nonlinear. A continuous nonlinear vector function $\boldsymbol{f}$ describes the dynamics of the state variable $\boldsymbol{z}$, with dimension $n$.

\subsection*{Concentrated Model of the DBMD} 
The topology of the circuit proposed by Hansen \etal\cite{hansen_double_2015} has been used as a starting point for the electrical representation of the concentrated model. There, a model based on experimental results was developed initially. In contrast to~\cite{hansen_double_2015} we have modified the components in the equivalent circuit with enhanced functionalities representing physical properties.

\begin{figure}[!ht]
	\centering
	\includegraphics[scale=0.9]{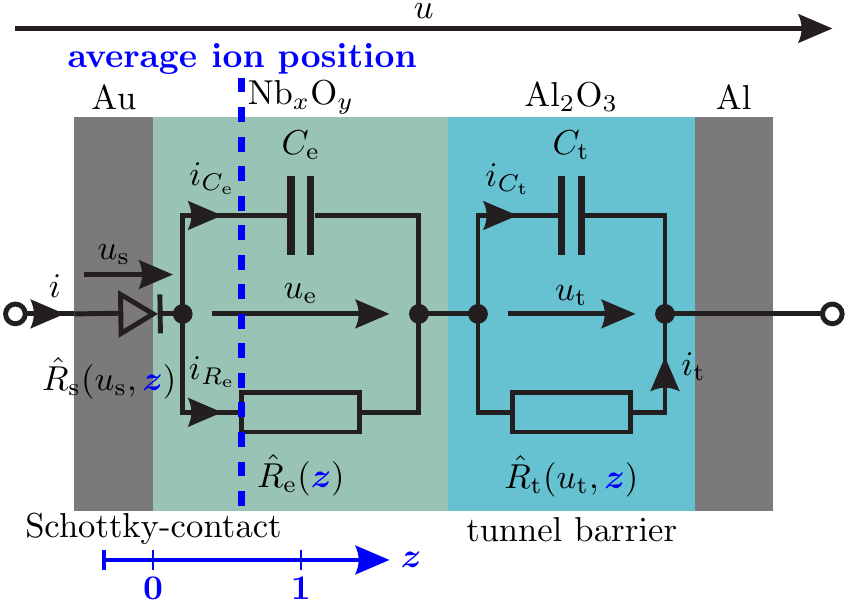}%
	\caption{Electrical representation of the device using state-dependent components and concentrated parameters.}%
	\label{fig.:electricalModel}%
\end{figure}

For the concentrated model, a deeper modeling approach depending on already known as well as novel physical insights is desired. On the other hand, the underlying physical and chemical phenomena are complex. Concerning the concentrated model, a preferably straightforward implementation of the functionality, without sacrificing the physical interpretation, is beneficial. Furthermore, it is advised to reformulate the mathematical description of the device in order to classify it as a memristive system. To this end, we focus the perspective rather on an electrical point of view than on a physical one by using normalized parameters and ordinary differential equations. This leads to the concentrated model depicted in figure~\ref{fig.:electricalModel}, where the Schottky-contact is modeled by a diode, and the electrolyte region and the tunnel barrier are represented each by a parallel connection of a resistor and a parasitic capacitor.
However, the resulting parameters and equations, which are electrically interpretable,  still have a corresponding physical meaning. In the following, physically motivated parameters and equations are deduced.

\subsection*{Memristive Behavior}
As a first step, we derive a concentrated formulation for the memristive behavior including the internal state, cf. equation~(\ref{eqn:memristiveSystemDz}). Physically, this internal state is represented by the distribution of ions in the electrolyte, which determines a total resistance value~\cite{hansen_double_2015,dirkmann_role_2016}. Individual regions as Schottky-barrier, solid state electrolyte and tunnel barrier are coupled through this internal state. In the distributed model the position of each single ion affects the overall resistance inherently. Instead, in the concentrated model, the ion distribution is expressed by the average ion position, see figure~\ref{fig.:electricalModel}, or more precisely, by the center of gravity of oxygen ions $x$. The internal state of the concentrated model is thus defined by a normalized average ion position
\begin{equation}
z = \frac{x-x_\mathrm{min}}{x_\mathrm{max}-x_\mathrm{min}}\:,\quad\mathrm{with}\quad z\in[0,1]\:,
\label{eq:normalizedStateVAriable}
\end{equation}
where $x_\mathrm{min}$ and $x_\mathrm{max}$ are the minimal and maximal absolute average positions, see figure~\ref{fig.:electricalModel} (blue, dashed line). Thus, $z=0$ corresponds to the low Ohmic state, whereas $z=1$ is related to the thermodynamical equilibrium or high Ohmic state. The thermodynamical equilibrium is given for homogeneously distributed positive and negative ions. This state is represented by an average ion position right in the middle of the electrolyte. With this, we set $x_\mathrm{min}=0$ and $x_\mathrm{max}=d_\mathrm{e}/2$, where $d_\mathrm{e}$ is the electrolyte width.

Inspired from the results of Meyer \etal\cite{meyer_oxide_2008}, we describe the motion of ions by an ion hopping phenomenon in terms of a jump over a potential barrier $\Phi_\mathrm{a}$, namely activation energy. It is a material property, which together with equation~(\ref{eqn:arrheniusLaw}) leads to an overall ion drift velocity. This drift velocity depends on the applied electrical field. In the concentrated model the electrical field is transformed to a corresponding voltage. Due to this transformation, nonlinearities between electrical fields and resulting voltages are neglected. With this, we can describe the normalized drift velocity of ions
\begin{equation}
\dot{z} = \frac{-\dot{Z}\,w(z)}{e^{\varphi_\mathrm{a}(u,z)}}\sinh\left(\frac{\,u_\mathrm{r}(u,u_\mathrm{s},z)+u_\mathrm{e}- U_\mathrm{c}}{U_\mathrm{e}}\right)
\label{eqn:memristiveBehavior}
\end{equation}
in terms of voltages instead of electrical fields. This velocity in turn is an ordinary differential equation describing the memristive behavior, cf. equation~(\ref{eqn:memristiveSystemDz}). The drift amplitude
\begin{equation}
\dot{Z} = 2\,\nu\,a\:,\quad\mathrm{with}\quad a=\frac{d_\mathrm{hop}}{x_\mathrm{max}-x_\mathrm{min}}
\end{equation}
contains the hopping frequency $\nu$ and a normalized hopping distance $a$, where $d_\mathrm{hop}$ denotes the absolute hopping distance per jump. 

A sum of voltages causes the ion motion, see equation~(\ref{eqn:memristiveBehavior}) and figure~\ref{fig.:electricalModel}. Contributing voltages are the electrolyte voltage $u_\mathrm{e}$, a state-dependent amount of the voltage drop over the Schottky-barrier $u_\mathrm{r}(u,u_\mathrm{s},z)$, with Schottky-barrier voltage $u_\mathrm{s}$ and a voltage $U_\mathrm{C}$ representing the Coulomb potential. The reference electrolyte voltage is defined as follows:
\begin{equation}
U_\mathrm{e} = \frac{2}{c}\frac{d_\mathrm{e}}{d_\mathrm{hop}}U_\mathrm{\vartheta}\:,
\quad\mathrm{with}\quad
U_\vartheta
= \frac{k_\mathrm{B}\,\vartheta}{q_\mathrm{e}}\:.
\end{equation}	
Here, $c$ is the charge number of the used material (Nb$_x$O$_y	$), $U_\vartheta$ is the thermal voltage and $q_\mathrm{e}$ is the elementary charge.

For the reset process, a particular contribution to the sum of voltages affecting the ion motion is given by an amount of the Schottky-voltage during the reset process. This amount increases especially for ions allocated next to this interface. Therefore a state as well as voltage-dependent function modeling this amount of contribution is needed. To this end, we define a function
\begin{equation}
u_\mathrm{r}(u,u_\mathrm{s},z) = \sigma(-u)\left[1-z\right]\,u_\mathrm{s}\:,
\label{eqn:schottkyVoltageAmount}
\end{equation}
where the dependency with respect to the state variable is assumed to be linear. Ions located directly at the interface, $z=0$, are affected by the total amount of the Schottky-barrier voltage, whereas for $z=1$ the amount of the Schottky-barrier voltage influencing the ion movement is zero. In equation~(\ref{eqn:schottkyVoltageAmount}), the unit step function
\begin{equation}
\sigma(\xi) = \left\{ \begin{array}{rcl}
1 & \mathrm{for} & \xi > 0\\
0 & \mathrm{otherwise} &
\end{array} \right.
\end{equation}
ensures that this amount of voltage applies only during the reset process, i.e. for $u<0$.

It is common practice to restrict the normalized state variable between 0 and 1 by using a window function, as shown by Biolek \etal\cite{biolek_reliable_2013}. For our approach, we need a modified window function, which overcomes the boundary lock problem. Preparing the window function of~\cite{biolek_reliable_2013} leads to an enhancement
\begin{equation}
w(z) = \left[1-2\,w_0\right]\left[1-\left[2\,z-1\right]^{2p}\right]+w_0\:,
\end{equation}
where the offset $w_0$ is positive and can be chosen arbitrarily small. This window function still involves the boundary caused nonlinearities, where $p$ controls the edge steepness.

Adsorption and desorption of ions at the Schottky-interface are further results from investigations with the distributed model~\cite{dirkmann_role_2016}. Besides, from long term investigations within the scope of this work, we assume that the number of defects is not constant. For a consistent electrical model, these effects must be considered. Due to the charge separation during the set process, the activation energy increases for $z\rightarrow 0$. This is because more and more ions are impounded at this interface for positive voltages. In addition, also defect formation is a slow process compared to ionic motion. On the other hand, a desorption mechanism becomes valid for negative applied voltages and reduces the velocity during the reset process. Since these processes act on different time scales but have to be taken into account within the concentrated model, a state-dependent activation energy is preferable. To this end, we introduce a state-dependent, normalized energy barrier
\begin{equation}
\varphi_\mathrm{a}(u,z) = \sigma(u)\left[\varphi_{\mathrm{a}_1}+z\left[\varphi_{\mathrm{a}_0}-\varphi_{\mathrm{a}_1}\right]-\varphi_{\mathrm{a_r}}\right]+\varphi_{\mathrm{a_r}}\:,
\label{eqn:stateDependentActivationEnergy}
\end{equation}
with $\Phi_\mathrm{a}=\varphi_\mathrm{a}\,k_\mathrm{B}\,\vartheta$. The indices $\lbrace0,1\rbrace$ denote the boundary values corresponding to the state, whereas $\varphi_{\mathrm{a_r}}$ is the assumed activation energy for the reset process. The linear dependency with respect to the state variable can be adapted by other functions considering different material properties. 

Beside the ordinary differential equation for the memristive behavior~(\ref{eqn:memristiveSystemDz}), we have to formulate expressions for the algebraic input-output relation~(\ref{eqn:memristiveSystemU}) between voltage and current. The derivation of these relations is shown in the next subsection.

\subsection*{Input Output Relations}
The DBMD consists of three state-dependent contiguous regions, see figure~\ref{fig.:electricalModel}. The regions can be interpreted as single memristive systems, all of them coupled by the state variable. In order to classify this device as a memristive system, we have to describe the input-output relations for each region like in equation~(\ref{eqn:memristiveSystemU}). In the sequel, the individual regions are distinguished.

\subsubsection*{Schottky-contact}
The Schottky-current can be described by the Schottky-equation. Hansen \etal\cite{hansen_double_2015} has used an additional fitting parameter for the reverse Schottky-current to fit the measured data.

Taking a barrier lowering due to image charges~\cite{sze_physics_2007} into account in combination with a state-dependent normalized Schottky-barrier height $\varphi_\mathrm{s}(z)$ and an ideality factor $n(z)$, the Schottky-current is given as
\begin{equation}
i_\mathrm{s}(u_\mathrm{s},z) = I_\mathrm{s}\,\rme^{-\left[\varphi_\mathrm{s}(z)+\alpha_\mathrm{f}\sqrt{\frac{|u_\mathrm{s}|-u_\mathrm{s}}{\alpha_\mathrm{s}\,U_\vartheta}}\right]}\left[\rme^{\frac{1}{n(z)}\frac{u_\mathrm{s}}{U_\vartheta}}-1\right]\:.
\label{eqn:schottkyEquation}
\end{equation}
Here, a dimensionless fitting parameter $\alpha_\mathrm{f}$ weights the barrier lowering term, which is caused by the Schottky-effect. The normalized Schottky-barrier thickness is
\begin{equation}
\alpha_\mathrm{s} = \frac{2\,d_\mathrm{s}}{D_\mathrm{s}}\:,
\quad\mathrm{with}\quad
D_\mathrm{s}
=\frac{q^2_\mathrm{e}}{4\pi\,\epsilon_0\,\epsilon_\mathrm{r}\,k_\mathrm{B}\,\vartheta}\:,
\end{equation}
where $d_\mathrm{s}$ denotes the absolute Schottky-barrier thickness and $D_\mathrm{s}$ is the normalization factor, with permittivity of vacuum $\epsilon_0$ and relative permittivity $\epsilon_\mathrm{r}$ of the electrolyte. The amplitude $I_\mathrm{s}=R_\mathrm{i}\,A\,\vartheta^2$ contains besides the temperature the effective Richardson constant $R_\mathrm{i}$ and the cross-sectional area $A$ of the device. Linear functions for the state-dependencies of the normalized Schottky-barrier height and the ideality factor have been assumed and can be expressed by
\begin{eqnarray}
\varphi_\mathrm{s}(z) = \varphi_{\mathrm{s}_0}+z\left[\varphi_{\mathrm{s}_1}-\varphi_{\mathrm{s}_0}\right]
\quad\mathrm{and}\\
n(z) = n_0+z\left[n_1-n_0\right]\:,
\end{eqnarray}
with $\varphi_{\mathrm{s_0}}\leq\varphi_\mathrm{s}(z)\leq\varphi_{\mathrm{s_0}}$ and $n_0\leq n(z)\leq n_1$. The resulting input-output relation between current through and voltage drop over the Schottky barrier is given by equation~(\ref{eqn:schottkyEquation}); it can be rewritten into the form of equation~(\ref{eqn:memristiveSystemU}) as
\begin{equation}
u_\mathrm{s} = \hat{R}_\mathrm{s}(u_\mathrm{s},z)\,i_\mathrm{s}(u_\mathrm{s},z)\:,
\end{equation}
thus defining resistance $\hat{R}_\mathrm{s}$. With this, we can identify the normal form of a memristive system in term of equations~(\ref{eqn:memristiveSystemU}) and (\ref{eqn:memristiveSystemDz}) for the Schottky-region.

\subsubsection*{Electrolyte region}
Memristive behavior of the electrolyte region stems from the actual ion distribution within the electrolyte, which influences the conductivity~\cite{waser_redox-based_2009}. A deeper modeling approach requires a state-dependent resistance for this region. For this purpose, we assume a linear state-dependency for the electrolyte region
\begin{equation}
\hat{R}_\mathrm{e}(z) = R_{\mathrm{e}_0}+z\left[R_{\mathrm{e}_1}-R_{\mathrm{e}_0}\right]\:,
\end{equation}
with $R_{\mathrm{e_0}}<\hat{R_\mathrm{e}}(z)<R_{\mathrm{e_1}}$ as the high and low resistance states, respectively. This in turn leads to the memristive normal form of equation~(\ref{eqn:memristiveSystemU}) and~(\ref{eqn:memristiveSystemDz}) regarding the input-output relation
\begin{equation}
u_\mathrm{e} = \hat{R}_\mathrm{e}(z)\,i_{R_\mathrm{e}}\:.
\end{equation}

\subsubsection*{Tunnel barrier}
The last region to address is the Al$_2$O$_3$ layer. This layer is an electrically high quality tunnel barrier~\cite{hansen_double_2015}. An electrical formulation of the tunnel current was given by Simmons~\cite{simmons_generalized_1963}. There, a distinction depending on the applied voltage was introduced: low voltage regime $|u_\mathrm{t}|\approx0$, intermediate regime $|u_\mathrm{t}|\leq\Phi_\mathrm{t}/q_\mathrm{e}$ and high voltage regime $|u_\mathrm{t}|>\Phi_\mathrm{t}/q_\mathrm{e}$, where $\Phi_\mathrm{t}$ denotes the tunnel barrier height and $u_\mathrm{t}$ the voltage drop over the tunnel barrier. Choosing the right equation for an accurate functionality of the concentrated model is important.

Kinetic Monte-Carlo simulations in~\cite{dirkmann_role_2016} have shown that the restriction $u_\mathrm{t}<3~\mathrm{V}$ is fulfilled for the considered operating range $-2~\mathrm{V}\leq u \leq 3~\mathrm{V}$ of the device. This justifies the use of the intermediate Simmons equation for the concentrated model, because the tunnel barrier height of the DBMD is about $\Phi_\mathrm{t}\approx3~\mathrm{eV}$, cf.~\cite{dirkmann_role_2016}. A concentrated formulation of the Simmons equation leads to the tunnel current
\begin{equation}
\eqalign{
i_\mathrm{t}(u_\mathrm{t},z) =  I_\mathrm{t}\frac{g(-u_\mathrm{t},z)-g(u_\mathrm{t},z)}{\alpha^2_\mathrm{t}(z)}\:,
\quad\mathrm{with}\quad\\
g(u_\mathrm{t},z) = \varphi_\mathrm{t}(u_\mathrm{t})\,\rme^{-\alpha_\mathrm{t}(z)\sqrt{\varphi_\mathrm{t}(u_\mathrm{t})}}\:,
\quad\mathrm{where}\\
\varphi_\mathrm{t} = \varphi_{\mathrm{t}_0}+\frac{1}{2}\frac{u_\mathrm{t}}{U_\vartheta}\:
\quad\mathrm{and}\quad
I_\mathrm{t}
= \frac{A}{D^2_\mathrm{t}}\frac{k_\mathrm{B}\,q_\mathrm{e}}{2\pi\,\mathrm{h}}\vartheta\:.
}
\end{equation}
The normalized tunnel barrier height is denoted by $\varphi_{t_0}$. Here, $\alpha_\mathrm{t}$ is the normalized tunnel barrier thickness, where the absolute width $d_\mathrm{t}$ is normalized by $D_\mathrm{t}=\frac{\mathrm{h}}{4\pi\sqrt{2\,m_\mathrm{e}\,k_\mathrm{B}\,\vartheta}}$. The normalization constant is based on physical constants like Planck's constant $h$ and mass of an electron $m_\mathrm{e}$.

Due to the Simmons formula~\cite{simmons_generalized_1963}, which accounts only for elastic tunneling, the effective barrier thickness depends on the ion distribution and with this on the internal state. This results in a state-dependent normalized tunnel barrier thickness. For the sake of an effective implementation, we have used a linear dependency, which leads to
\begin{equation}
\alpha_\mathrm{t}(z) = \alpha_{\mathrm{t}_0}+z\left[\alpha_{\mathrm{t}_1}-\alpha_{\mathrm{t}_0}\right]\:,
\:\mathrm{with}\:
\alpha_{\mathrm{t_0}}\leq \alpha(z)\leq\alpha_{\mathrm{t_1}}\:.
\end{equation}

As it can be seen, the input-output relation of the tunnel barrier region reads 
\begin{equation}
u_\mathrm{t} = \hat{R}_\mathrm{t}(u_\mathrm{t},z)\,i_\mathrm{t}(u_\mathrm{t},z)\:,
\end{equation}
which is a memristive system, cf. equations~(\ref{eqn:memristiveSystemU}) and~(\ref{eqn:memristiveSystemDz}).

\section{Simulation Results of the Concentrated Model}
Several simulation models of memristive systems are available in the literature~\cite{biolek_reliable_2013}. Most of them are based on mathematical descriptions, where the physical meanings of parameters are not obvious.

In this section, we want to verify the concentrated model by comparisons with kinetic Monte-Carlo simulations as well as measurements. To this end, an LTSpice implementation of the concentrated model is used. For investigations of small time scales $T\approx100\,\mathrm{s}$ hysteresis curves are utilized. In contrast to that, long time-scales $T\approx600\,\mathrm{s}$ are investigated by the step response of the device for two amplitudes. A semilogarithmic scale with respect to the absolute value of the current is used for the hysteresis curves. This improves the visibility of a gradual resistance change. To be consistent with measurements and kinetic Monte-Carlo simulations, the high resistance state $z=1$ is chosen as the initial state for all following simulations.

\subsection*{Short Time Scale Simulations: Hysteresis Curves}
\begin{figure*}[!hbt]
	\centering
	\includegraphics[scale=0.55]{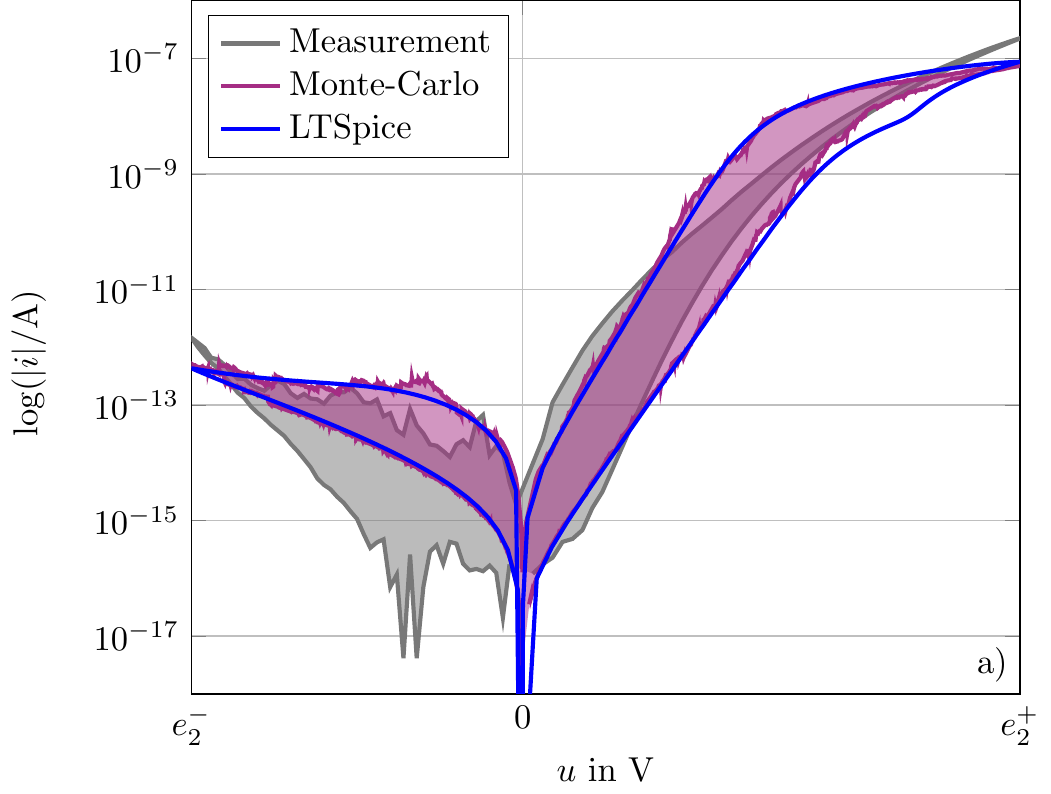}\quad%
	\includegraphics[scale=0.55]{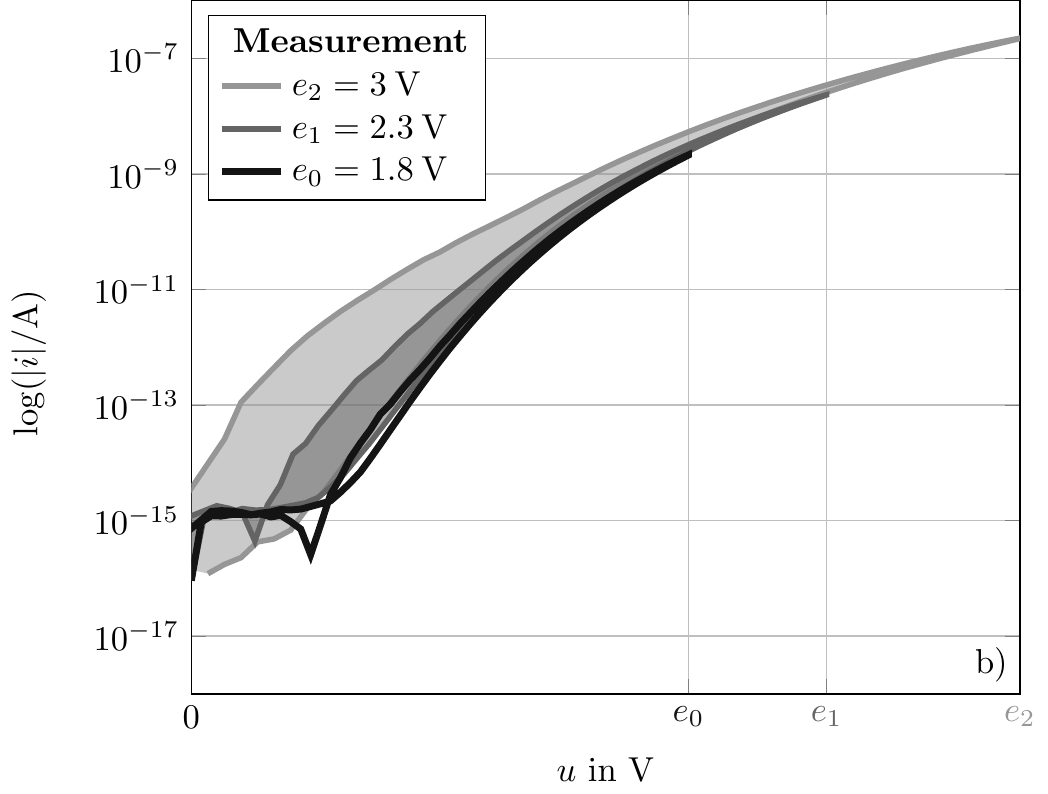}\\[0.5cm]%
	\includegraphics[scale=0.55]{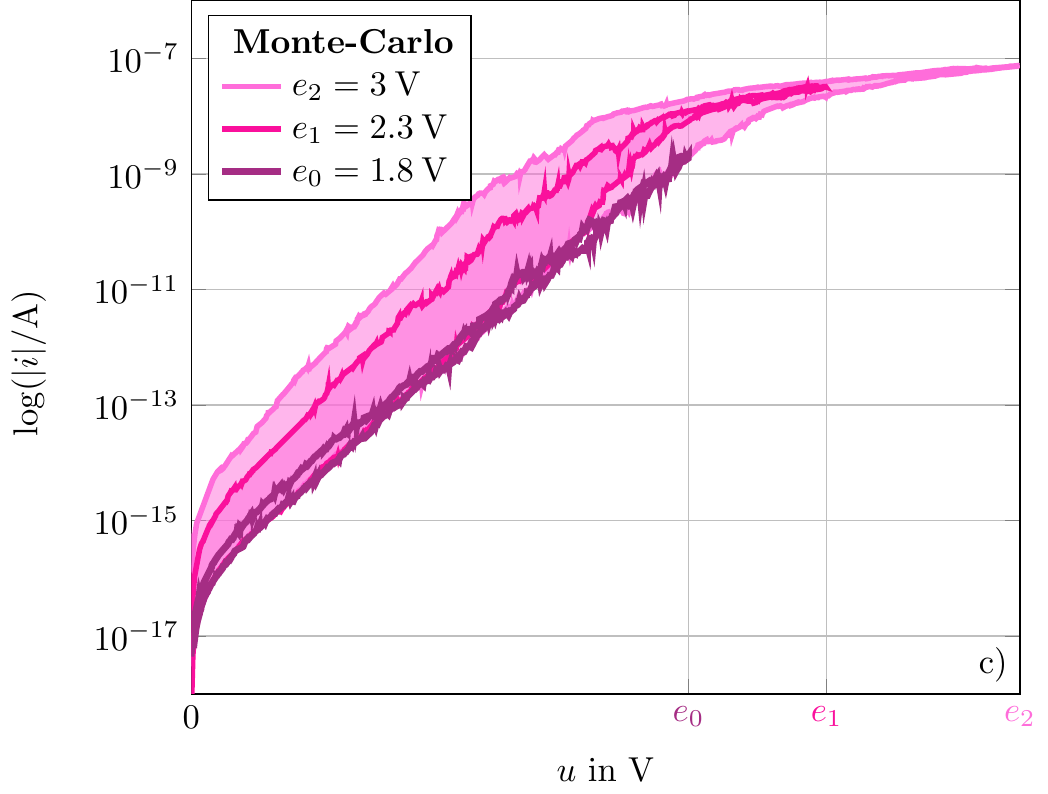}\quad%
	\includegraphics[scale=0.55]{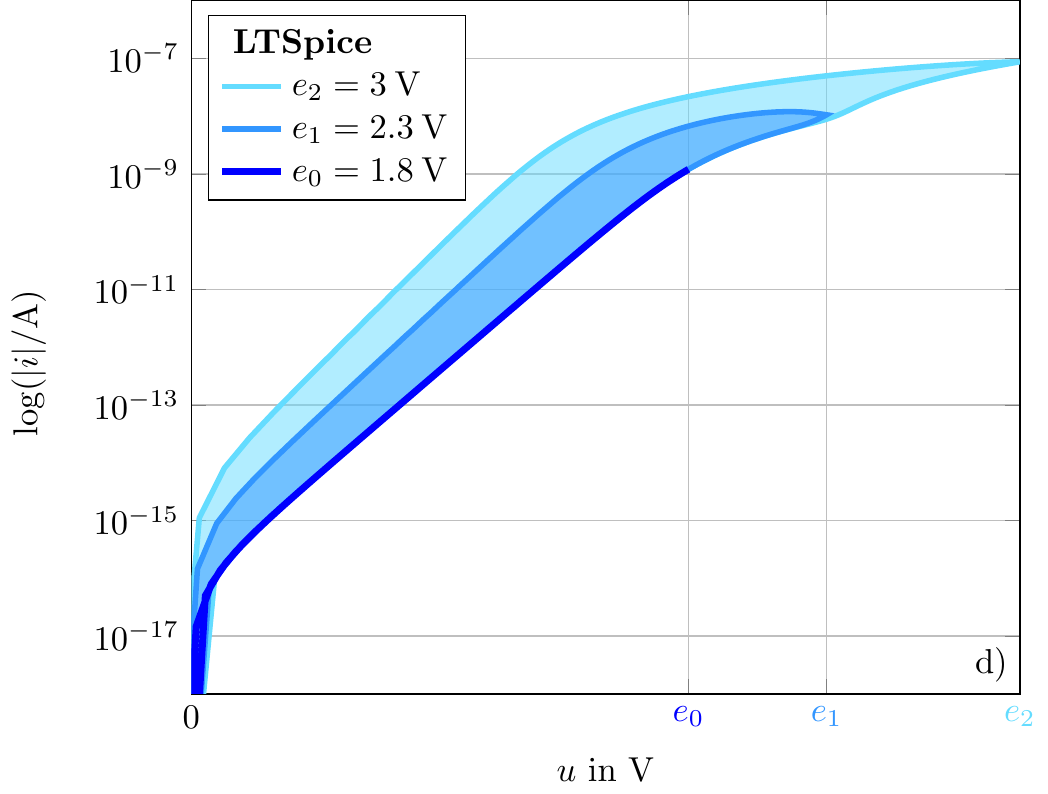}%
	\caption{a) Comparison of hysteresis curves for a triangle input voltage (inset figure~\ref{fig.:measuredIV}) with $e_2^+=3~\mathrm{V}$ and $e_2^-=-2~\mathrm{V}$. b)-d) Hysteresis curves for peak voltages $e_0=1.8~\mathrm{V}$, $e_1=2.3~\mathrm{V}$, $e_2=3~\mathrm{V}$: b) measured, c) distributed model, d) concentrated model.}%
	\label{fig.:hysteresis}%
\end{figure*}

Short time scale investigations are done by considering resulting hysteresis curves for input voltages of the form depicted in figure~\ref{fig.:measuredIV} (inset). For the sake of consistency, the amplitudes are chosen as in measurements and kinetic Monte-Carlo simulations. The results are shown in figure~\ref{fig.:hysteresis}, where for improved clarity the areas of some hysteresis loops have been shaded. In figure~\ref{fig.:hysteresis} a), hysteresis curves for exactly the same input voltage of figure~\ref{fig.:measuredIV} are shown. The good coincidence between LTSpice simulation, kinetic Monte-Carlo simulation and measurement is remarkable, regarding the complexity of the distributed model compared to the concentrated. 

\begin{figure}[!hbt]
	\centering
	\includegraphics[scale=0.75]{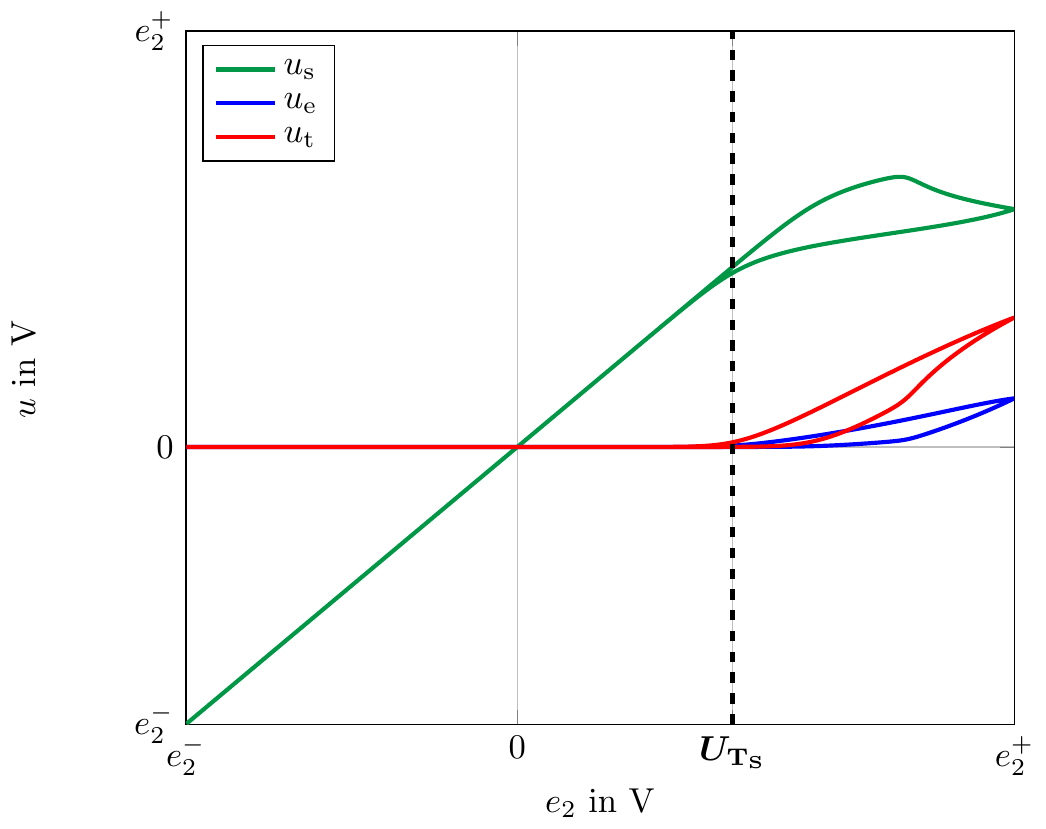}
	\caption{Voltage drop over different regions of the device with respect to the input voltage of figure~\ref{fig.:measuredIV}.}
	\label{fig:voltages}
\end{figure}

In figure~\ref{fig.:hysteresis} b)-d) hysteresis curves for different voltage amplitudes are depicted.
In coincidence with measurements and kinetic Monte-Carlo simulations, resulting curves for the concentrated model show that a threshold voltage $U_{\mathrm{T_S}}$ has to be reached to induce a memristive behavior, because for voltages below $U_{\mathrm{T_S}}$ no hysteresis occurs. To emphasize this behavior, figure~\ref{fig:voltages} shows the voltage drops over individual regions with respect to the total voltage drop over the device.
As expected, until $U_{\mathrm{T_S}}$ is reached almost the whole amount of the applied voltage drops over the Schottky-region. When the voltage reaches $U_{T_S}$, the Schottky-contact becomes more and more a short-circuit. The forward direction of the diode leads to an increasing voltage drop over the electrolyte as well as over the tunnel barrier. The voltage drop over the electrolyte results in a change of the state variable and therefore in a decreasing of the device resistance, cf.~\cite{dirkmann_role_2016}.  

\subsection*{Long Time Scale Simulations: Step Responses}
\begin{figure*}[!ht]
	\centering
	\includegraphics[scale=0.55]{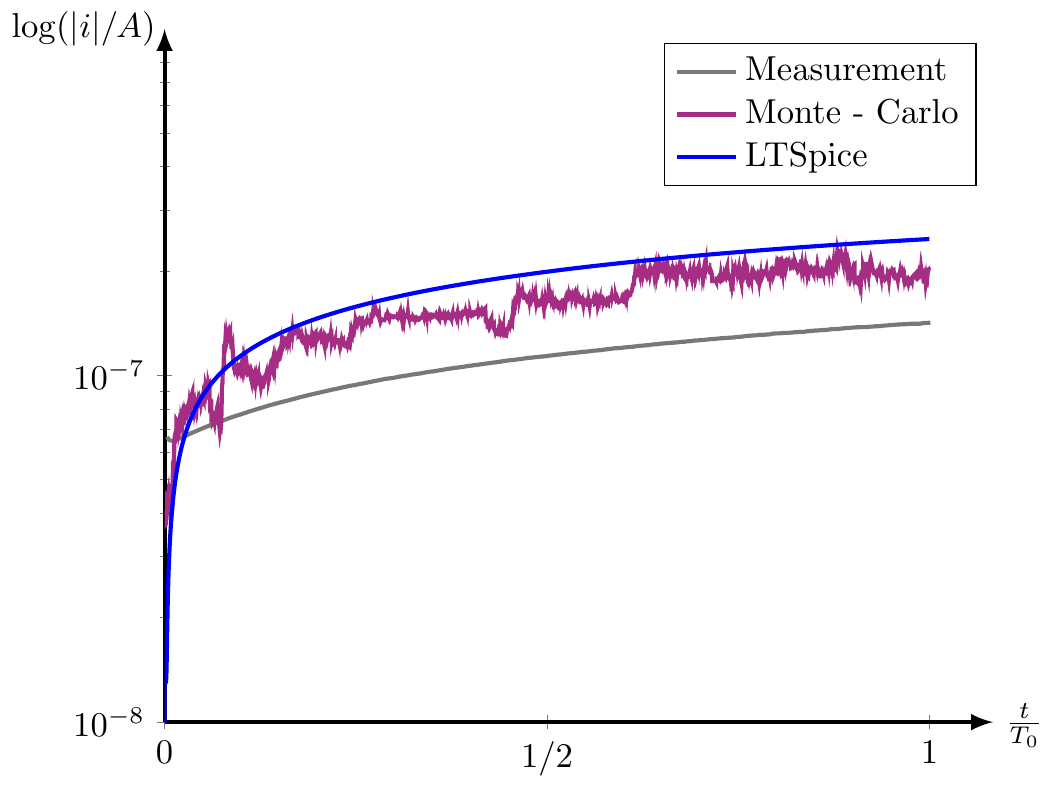}\quad%
	\includegraphics[scale=0.55]{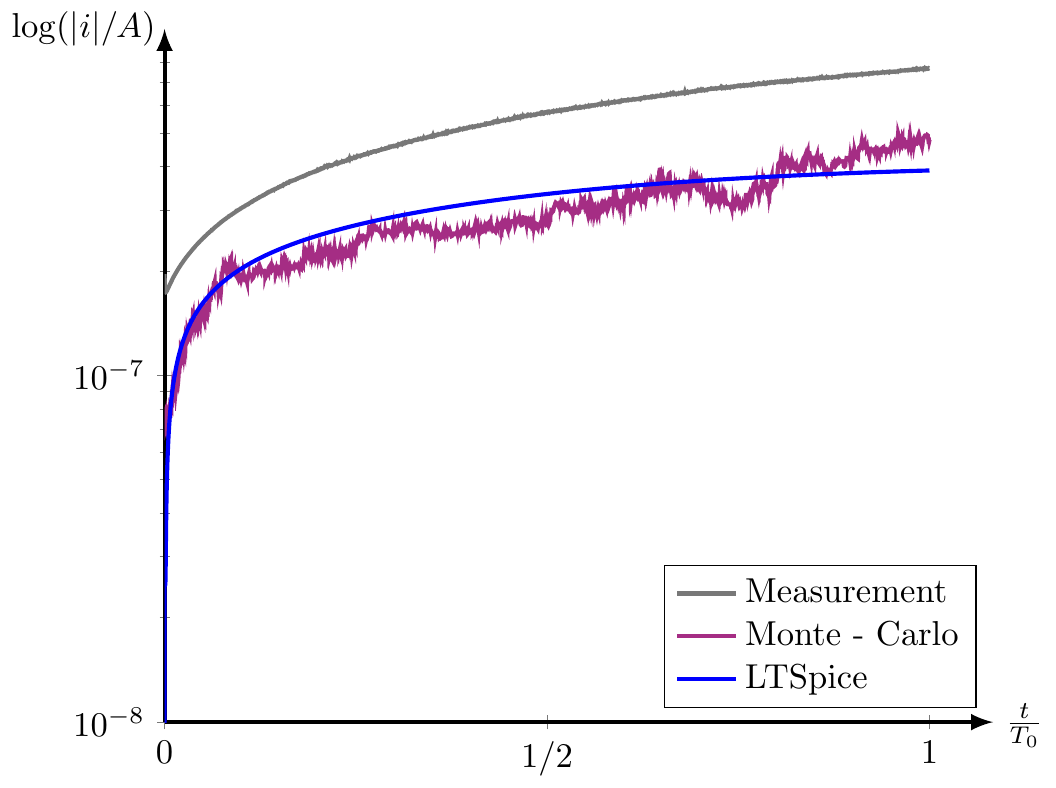}%
	\caption{Measured and simulated step responses for input voltage amplitudes of $2.5~\mathrm{V}$ (left) and $2.9~\mathrm{V}$ (right). The measurement and simulation time is $T_0=600\:\mathrm{s}$}%
	\label{fig.:Step2_5}%
\end{figure*}

For long time scale investigations, step responses of the device, with two different voltage amplitudes, have been measured and simulated. In figure~\ref{fig.:Step2_5}, the step response for an input voltage amplitude of $2.5~\mathrm{V}$ (left) and $2.9~\mathrm{V}$ (right) is presented.
Again a good coincidence between concentrated and distributed model can be seen. This is comprehensible since the concentrated model is based on the distributed model. However, it should be stressed that further investigations, like the step response, were initiated by the concentrated approach. On top of that, the agreement with measured data is also important.

We see from simulation results, that the current is higher for an input voltage amplitude of $2.5~\mathrm{V}$ and lower for an input voltage amplitude of $2.9~\mathrm{V}$ compared to the measurement. Similar discrepancies can be observed in the hysteresis curve of figure~\ref{fig.:hysteresis} a). We have to mention that the model has of course an imprecision which could lead to such discrepancies.

\section*{Conclusion}
In this work, we derived a concentrated model of a double barrier memristive device based on investigations of a distributed model and measurements. The concentrated model increases the mathematical accessibility, while maintaining physically meaningful parameters. 

The DBMD has been classified as a memristive system, with the average ion position as its state, i.~e. the drift velocity represents the memristive behavior. In this memristive system, a modified window function overcomes the boundary lock problem. For a gradual resistance change, physically justified state-dependencies of particular parameters have been utilized. Novel insights from actual research results, like adsorption and desorption mechanisms or ion formations within the electrolyte for long time scales, have been incorporated by an appropriate activation energy depending on both the state variable and the applied voltage. It is notable, that the presented approach can also be used for physical investigations. The step response of the DBMD was initiated during the modeling procedure of the concentrated model.     

Although the distributed model is more complex than the concentrated one, LTSpice simulation results have shown a good coincidence compared with kinetic Monte-Carlo simulations as well as measurements. As an example, the simulation time with LTSpice was on the time scale of seconds, whereas a kinetic Monte-Carlo simulation with a distributed model takes approximately some hours. 

This general approach in combination with physically meaningful parameters restrict the presented model not to a particular device. Instead, other resistive switching devices with different material compositions can be modeled only by adapting corresponding parameters. In conclusion, the concentrated model offers new possibilities for the investigation of complex neuromorphic circuits including real memristive devices, e.g. sensitivity analyses with respect to noisy conditions and parameter spread.

\section*{Acknowledgements}
The financial support by the German Research Foundation (Deutsche Forschungsgemeinschaft - DFG) through FOR 2093 is gratefully acknowledged.
\section*{References}
\bibliography{bibliography}

\end{document}